\documentclass{sig-alternate}

\usepackage{amssymb}
\usepackage{graphicx}
\usepackage{url}
\usepackage{booktabs}
\usepackage{ctable}
\usepackage{color}
\usepackage{xcolor}
\usepackage{algorithmic}
\usepackage{algorithm}
\usepackage{colortbl}
\usepackage{array}
\usepackage{appendix}

\newcommand{\Acal}{\ensuremath{\mathcal{A}}}
\newcommand{\Ocal}{\ensuremath{\mathcal{O}}}
\newcommand{\Qcal}{\ensuremath{\mathcal{Q}}}
\newcommand{\Rcal}{\ensuremath{\mathcal{R}}}
\newcommand{\Ucal}{\ensuremath{\mathcal{U}}}
\newcommand{\qbf}{\ensuremath{\mathbf{q}}}
\newcommand{\sbf}{\ensuremath{\mathbf{s}}}

\definecolor{darkviolet}{RGB}{148,0,211}

\begin{document}


\title{An IR-based Evaluation Framework for\\Web Search Query Segmentation}

\numberofauthors{2}
\author{
\alignauthor
Rishiraj Saha Roy and Niloy Ganguly\\
       \affaddr{Indian Institute of Technology Kharagpur}\\
       \affaddr{Kharagpur, West Bengal, India - 721302.}\\
       \email{\{rishiraj, niloy\}@cse.iitkgp.ernet.in}
\alignauthor
Monojit Choudhury and\\Srivatsan Laxman\\
       \affaddr{Microsoft Research India}\\
       \affaddr{Bangalore, Karnataka, India - 560025.}\\
       \email{\{monojitc, slaxman\}@microsoft.com}
}

\maketitle
\begin{abstract}
This paper presents the first evaluation framework for Web search query segmentation based directly on IR performance. In the past, segmentation strategies were mainly validated against manual annotations. Our work shows that the goodness of a segmentation algorithm as judged through evaluation against a handful of human annotated segmentations hardly reflects its effectiveness in an IR-based setup. In fact, state-of the-art algorithms are shown to perform as good as, and sometimes even better than human annotations -- a fact masked by previous validations. The proposed framework also provides us an objective understanding of the gap between the present best and the best possible segmentation algorithm. We draw these conclusions based on an extensive evaluation of six segmentation strategies, including three most recent algorithms, vis-\`{a}-vis segmentations from three human annotators. The evaluation framework also gives insights about which segments should be necessarily detected by an algorithm for achieving the best retrieval results. The meticulously constructed dataset used in our experiments has been made public for use by the research community.
\end{abstract}




\section{Introduction}
\label{sec:introduction}

Query segmentation is the process of dividing a query into individual semantic units~\cite{bergsma:07}. For example, the query {\tt singular value decomposition online demo} can be broken into {\tt singular value decomposition} and {\tt online demo}. All documents containing the individual terms {\tt singular}, {\tt value} and {\tt decomposition} are not necessarily relevant for this query. Rather, one can almost always expect to find the segment {\tt singular value decomposition} in the relevant documents. In contrast, although {\tt online demo} is a segment, finding the phrase or some variant of it may not affect the relevance of the document. Hence, the potential of query segmentation goes beyond the detection of multiword named entities. Rather, segmentation leads to a better understanding of the query and is crucial to the search engine for improving Information Retrieval (IR) performance. 

There is broad consensus in the literature that query segmentation can lead to better retrieval performance~\cite{bendersky:09,bergsma:07,hagen:11,li:11,tan:08}. 
However, most automatic segmentation techniques \cite{bergsma:07,brenes:10,hagen:11,li:11,tan:08,zhang:09} have so far been evaluated only against a small set of $500$ queries segmented by human annotators. 
Such an approach implicitly assumes that a segmentation technique that scores better against human annotations will also automatically lead to better IR performance. We challenge this approach on multiple counts. First, there has been no systematic study that establishes the quality of human segmentations in the context of IR performance. Second, grammatical structure in queries is not as well-understood as natural language sentences where human annotations have proved useful for training and testing of various Natural Language Processing (NLP) tools. This leads to considerable inter-annotator disagreement when humans segment search queries. Third, good quality human annotations for segmentation can be difficult and expensive to obtain for a large set of test queries. Thus, there is a need for a more direct IR-based evaluation framework for assessing query segmentation algorithms. This is the central motivation of the present work. 

We propose an IR-based evaluation framework for query segmentation that requires only human relevance judgments (RJs) for query-URL pairs for computing the performance of a segmentation algorithm -- such relevance judgments are anyway needed for training and testing of any IR engine. A fundamental problem in designing an IR-based evaluation framework for segmentation algorithms is to decouple the effect of segmentation accuracy from the way segmentation is used for IR. This is because a query segmentation algorithm breaks the input query into, typically, a non-overlapping sequence of words (segments), but it does not prescribe how these segments should be used during the retrieval and ranking of the documents for that query. We resolve this problem by providing a formal model of query expansion for a given segmentation; the various queries obtained can then be issued to any standard IR engine, which we assume to be a black box.

We conduct extensive experiments within our framework to understand the performance of several state-of-the-art query segmentation schemes~\cite{hagen:11,li:11,mishra:11} and segmentations by three human annotators. Our experiments reveal several interesting facts such as: (a) Segmentation is actively useful in improving IR performance, even though submitting all segments (detected by an algorithm) in double quotes to the IR engine degrades performance; (b) All segmentation strategies, including human segmentations, are yet to reach the best achievable limits in IR performance; (c) In terms of IR metrics, some of the segmentation algorithms perform as good as the best human annotator and better than the average/worst human annotator; (d) Current match-based metrics for comparing query segmentation against human annotations are only weakly correlated with the IR-based metrics, and cannot be used as a proxy for IR performance; and (e) There is scope for improvement for the matching metrics that compare segmentations against human annotations by differentially penalizing the straddling, splitting and joining of reference segments. In short, the proposed evaluation framework not only provides a formal way to compare segmentation algorithms and estimate their effectiveness in IR, but also helps us to understand the gaps in human annotation-based evaluation. The framework also provides valuable insights regarding the segmentations that can be used for improvement of the algorithms.


The rest of the paper is organized as follows. Sec.~\ref{sec:evaluation} introduces our evaluation framework and its design philosophy. Sec.~\ref{sec:experiments} presents the dataset and the segmentation algorithms compared on our framework. Sec.~\ref{sec:discussion} discusses the experimental results and insights derived from them.  In Sec.~\ref{sec:relatedis}, we discuss a few related issues, and the next section (Sec.~\ref{sec:relatedwk}) gives a brief background of past approaches to evaluate query segmentation and their limitations. We conclude by summarizing our contributions and suggesting future work in Sec.~\ref{sec:conclusions}.

\section{The evaluation framework}
\label{sec:evaluation}

In this section we present a framework for the evaluation of query segmentation algorithms based on IR performance. Let $\qbf$ denote a search query and let $\sbf^\qbf=\langle s^\qbf_1,\ldots,s^\qbf_n\rangle$ denote a segmentation of $\qbf$ such that a simple concatenation of the $n$ segments equals $\qbf$, i.e., we have $\qbf=(s^\qbf_1+\cdots+s^\qbf_{n})$, where + represents the concatenation operator. We are given a segmentation algorithm $\Acal$ and the task is to evaluate its performance. We require the following resources:

\begin{enumerate}
\item A test set $\Qcal$ of unquoted search queries.
\item A set $\Ucal$ of documents (or URLs) out of which search results will be retrieved.
\item Relevance judgments $r(\qbf,u)$ for query-URL pairs\\ $(\qbf,u)\in \Qcal\times\Ucal$. The set of all relevance judgments are collectively denoted by $\Rcal$.
\item An IR engine that supports quoted queries as input.
\end{enumerate}

The resources needed by our evaluation framework are essentially the same as those needed for the training and testing of a standard IR engine, namely, queries, a document corpus and set of relevance judgments. Akin to the training examples required for an IR engine, we only require relevance judgments for a small and appropriate subset of $\Qcal\times\Ucal$ (each query needs only the documents in its own pool to be judged)~\cite{voorhees:00}.

It is useful to separate the evaluation of segmentation performance, from the question of how to best exploit the segments to retrieve the most relevant documents. From an IR perspective, a natural interpretation of a segment could be that it consists of words that must appear together, in the same order, in documents where the segment is deemed to match~\cite{bergsma:07}. This can be referred to as \textit{ordered contiguity matching}. While this can be easily enforced in modern IR engines through use of double quotes around segments, we observe that not all segments must be used this way (see~\cite{metzler:05} for related ideas and experiments in a different context). Some segments may admit more general matching criteria, such as \textit{unordered or intruded contiguity} (e.g., a segment {\tt a b} may be allowed to match {\tt b a} or {\tt a c b} in the document). The case of unordered intruded matching may be restricted under \textit{linguistic dependence} assumptions (e.g., {\tt a b} can match {\tt a \textit{of} b} or {\tt b \textit{in} a}). Finally, some segments may even play non-matching roles (e.g., when the segment specifies user intent, like {\tt how to} and {\tt where is}). Thus, there may be several different ways to exploit the segments discovered by a segmentation algorithm. Even within the same query, different segments may need to be treated differently. For instance, in the query {\tt cannot view | word files | windows 7}, the first one might be matched using intruded ordered occurrence ({\tt cannot \textit{properly} view}), the second segment may be matched under a linguistic dependency model ({\tt files \textit{in} word}) and the last one under ordered contiguity. 

Intruded contiguity and linguistic dependency may be difficult to implement for the broad class of general Web search queries. Identifying how the various segments of a query should be ideally matched in the document is quite a challenging and unsolved research problem. On the other hand, an exhaustive expansion scheme, where every segment is expanded in every possible way, is computationally expensive and might introduce noise. Moreover, current commercial IR engines do not support any syntax to specify linguistic dependence or intruded or unordered occurrence based matching. Hence, in order to keep the evaluation framework in line with the current IR systems, we focus on ordered contiguity matching which is easily implemented through the use of double quotes around segments. However, we note that the philosophy of the framework does not change with increased sophistication in the retrieval system -- only the expansion sets for the queries have to be appropriately modified.

\begin{table}[t]
	\centering
	\caption{Example of generation of quoted versions for a segmented query.}
	\resizebox{0.45\textwidth}{!}{
		\begin{tabular} {c c}
		\\ \toprule
		\textbf{Segmented query} 								& \textbf{Quoted versions} 									\\ \toprule
																						& {\tt we are the people song lyrics} 			\\
																						& {\tt we are the people "song lyrics"} 		\\
																						& {\tt we are "the people" song lyrics} 		\\
		{\tt we are | the people | song lyrics}	& {\tt we are "the people" "song lyrics"} 	\\
																						& {\tt "we are" the people song lyrics} 		\\
																						& {\tt "we are" the people "song lyrics"} 	\\
																						& {\tt "we are" "the people" song lyrics} 	\\
																		 				& {\tt "we are" "the people" "song lyrics"}	\\  \bottomrule
		\end{tabular}}
	\label{tab:versions}
\end{table}

We propose an evaluation framework for segmentation algorithms that generates all possible quoted versions of a segmented query (see Table~\ref{tab:versions}) and submits each quoted version to the IR engine. The corresponding ranked lists of retrieved documents are then assessed against relevance judgments available for the query-URL pairs. The IR quality of the best-performing quoted version is used to measure performance of the segmentation algorithm. 
We now formally specify our evaluation framework that computes what we call a Quoted Version Retrieval Score (QVRS) for the segmentation algorithm given the test set $\Qcal$ of queries, the document pool $\Ucal$ and the relevance judgments $\Rcal$ for query-URL pairs.

\subsubsection*{Quoted query version generation}

Let the segmentation output by algorithm $\Acal$ be denoted by $\Acal(\qbf)=\sbf^\qbf=\langle s^\qbf_1,\ldots,s^\qbf_{n}\rangle$. We generate all possible {\em quoted versions} of the query $\qbf$ based on the segments in $\Acal(\qbf)$. In particular, we define $\Acal_0(\qbf)=(s^\qbf_1+\cdots+s^\qbf_{n})$ with no quotes on any of the segments, $\Acal_1(\qbf)=(s^\qbf_1+\cdots+\mathrm{``}s^\qbf_{n}\mathrm{"})$ with quotes only around the last segment $s^\qbf_{n}$, and so on. Since there are $n$ segments in $\Acal(\qbf)$, this process will generate $2^n$ versions of the query, $\Acal_i(\qbf)$, $i=0,\ldots, 2^n-1$. We note that if $b_i=(b_{i1},\ldots,b_{in})$ be the $n$-bit binary representation of $i$, then $\Acal_i(\qbf)$ will apply quotes to the $j^\mathrm{th}$ segment $s^\qbf_j$ iff $b_{ij}=1$. We deduplicate this set, because $\{\Acal_i(\qbf)\::\:i=0,\ldots,2^n-1\}$ can contain multiple versions that essentially represent the same quoted query version (when single words are inside quotes). For example, the query versions {\tt "harry potter" "game"} and {\tt "harry potter" game} are equivalent in terms of the input semantics of an IR engine. The resulting set of unique quoted query versions is denoted $\Qcal_\Acal(\qbf)$.

\subsubsection*{Document retrieval using IR engine}

For each $\Acal_i(\qbf)\in\Qcal_\Acal(\qbf)$ we use the IR engine to retrieve a ranked list $\Ocal_i$ of documents out of the document pool $\Ucal$ that matched the given quoted query version $\Acal_i(\qbf)$. The number of documents retrieved in each case depends on the IR metrics we will want to use to assess the quality of retrieval. For example, to compute an IR metric at the top $k$ positions, we would require that at least $k$ documents be retrieved from the pool.

\subsubsection*{Measuring retrieval against relevance judgments}

Since we have relevance judgments ($\Rcal$) for query-URL pairs in $\Qcal\times\Ucal$, we can now compute IR metrics such as normalized Discounted Cumulative Gain (nDCG), Mean Average Precision (MAP) or Mean Reciprocal Rank (MRR) to measure the quality of the retrieved ranked list $\Ocal_i$ for query $\qbf$. We use $@k$ variants of each of these measures which are defined to be the usual metrics computed after examining only the top-$k$ positions. For example, we can compute $\mathrm{nDCG@}k$ for query $\qbf$ and retrieved document-list $\Ocal_i$ using the following formula:
\begin{equation}
\mathrm{nDCG@}k(\qbf,\Ocal_i\:,\:\Rcal) = r(\qbf,\Ocal_i^1) + \sum_{j=2}^k \frac{r(\qbf,\Ocal_i^j)}{\log_2 j}
\end{equation}
where $\Ocal_i^j$, $j=1,\ldots,k$, denotes the $j^\mathrm{th}$ document in the ranked-list $\Ocal_i$ and $r(\qbf,\Ocal_i^j)$ denotes the associated relevance judgment from $\Rcal$.

\subsubsection*{Oracle score using best quoted query version}

Different quoted query versions $\Acal_i(\qbf)$ (all derived from the same basic segmentation $\Acal(\qbf)$ output by the segmentation algorithm $\Acal$) retrieve different ranked lists of documents $\Ocal_i$. As discussed earlier, automatic apriori selection of a good (or the best) quoted query version is a difficult problem. While different strategies may be used to select a quoted query version, we would like our evaluation of the segmentation algorithm $\Acal$ to be agnostic of the version-selection step. To this end, we select the best-performing $\Acal_i(\qbf)$ from the entire set $\Qcal_\Acal(\qbf)$ of query versions generated and use it to define our {\em oracle score} for $\qbf$ and $\Acal$ under the chosen IR metric~\cite{lease:09}. For example, the oracle score for nDCG@$k$ is as defined below:
\begin{equation}
\Omega_{\mathrm{nDCG@}k}(\qbf,\Acal) = \max_{\Acal_i(\qbf)\in\Qcal_\Acal(\qbf)} \mathrm{nDCG@}k(\qbf,\Ocal_i\:,\:\Rcal)
\label{eq:oracle}
\end{equation}
where $\Ocal_i$ denotes the ranked list of documents retrieved by the IR engine when presented with $\Acal_i(\qbf)$ as the input. We note that $\Qcal_\Acal(\qbf)$ always contains the original unsegmented version of the query. We refer to such an $\Omega_\cdot(\cdot,\cdot)$ as the \textit{Oracle}.



This forms the basis of our evaluation framework. We note that there can also be other ways to define this oracle score. For example, instead of seeking the best IR performance possible across the different query versions, we could also seek the minimum performance achievable by $\Acal$ irrespective of what version-selection strategy is adopted. This would give us a lower bound on the performance of the segmentation algorithm. However, the main drawback of this approach is that the minimum performance is almost always achieved by the fully quoted version (where every segment is in double quotes) (see Table~\ref{tab:bing_api}). Such a lower bound would not be useful in assessing the comparative performance of segmentation algorithms.

\subsubsection*{QVRS computation}

Once the oracle scores are obtained for all queries in the test set $\Qcal$, we can compute the average oracle score achieved by $\Acal$. We refer to this as the Quoted Version Retrieval Score (QVRS) of $\Acal$ with respect to test set $\Qcal$, document pool $\Ucal$ and relevance judgments $\Rcal$. For example, using the oracle with the nDCG@$k$ metric, we can define the QVRS score as follows:
\begin{equation}
QVRS(\Qcal,\Acal,{\mathrm{nDCG@}k})=\frac{1}{|\Qcal|}\sum_{\qbf\in\Qcal} \Omega_{\mathrm{nDCG@}k}(\qbf,\Acal)
\end{equation}
Similar QVRS scores can be computed using other IR metrics such as MAP@$k$ and MRR@$k$. In our experiments section, we report results using nDCG@$k$, MAP@$k$, and MRR@$k$, for $k=5$ and $k=10$ as most Web users examine only the first five or ten search results.

\section{Dataset and algorithms}
\label{sec:experiments}

In this section, we describe the dataset used and briefly introduce the algorithms compared on our framework. 

\subsection{Test set of queries ($\Qcal$)}
\label{subsec:querytestset}

We selected a random subset of $500$ queries from a slice of the query logs of Bing Australia\footnote{\scriptsize \url{http://www.bing.com/?cc=au}} containing $16.7$ million queries issued over a period of one month (May $2010$). We used the following criteria to filter the logs before extracting a random sample: (1) Exclude queries with non-ASCII characters, (2) Exclude queries that occurred fewer than 5 times in the logs (rarer queries often contained spelling errors), and (3) Restrict query lengths to between five and eight words. Shorter queries rarely contain multiple multiword segments, and when they do, they are mostly named entities that can be easily detected using dictionaries. Moreover, traditional search engines usually give satisfactory results for short queries. On the other hand, queries longer than eight words (only $3.24\%$ of all queries in our log) are usually error messages, complete NL sentences or song lyrics, that need to be addressed separately.

We denote this set of $500$ queries by $\Qcal$, the test set of unsegmented queries needed for all our evaluation experiments. The average length of queries in $\Qcal$ (our dataset) is $5.29$ words. The average query length was $4.31$ words in the Bergsma and Wang $2007$ Corpus\footnote{\scriptsize \url{http://bit.ly/xoyT2c}} (henceforth, BWC07)~\cite{bergsma:07}. Each of these $500$ queries were independently segmented by three human annotators (who issue around $20$-$30$ search queries per day) who were asked to mark a contiguous chunk of words in a query as a {\em segment} if they thought that these words together formed a coherent semantic unit. The annotators were free to refer to other resources and Web search engines during the annotation process, especially for understanding the query and its possible context(s). We shall refer to the three sets of annotations (and also the corresponding annotators) as $H_{A}$, $H_{B}$ and $H_{C}$.

It is important to mention that the queries in $\Qcal$ have some amount of word level overlap, even though all the queries have very distinct information needs. Thus, a document retrieved from the pool might exhibit good term level match for more than one query in $\Qcal$. This makes our corpus an interesting testbed for experimenting with different retrieval systems. There are existing datasets, including BWC07, that could have been used for this study. However, refer to Sec.~\ref{subsec:newdata} for an account of why building this new dataset was crucial for our research. 


\subsection{Document pool ($\Ucal$) and RJs ($\Rcal$)}
\label{subsec:docpool}

Each query in $\Qcal$ was segmented using all the nine segmentation strategies considered in our study (six algorithms and three humans). For every segmentation, all possible quoted versions were generated (total $4,746$) and then submitted to the Bing API\footnote{\scriptsize \url{http://msdn.microsoft.com/en-us/library/dd251056.aspx}} and the top ten documents were retrieved. We then deduplicated these URLs to obtain $14,171$ unique URLs, forming $\Ucal$. On an average, adding the $9^{th}$ strategy to a group of the remaining eight resulted in about one new quoted version for every two queries. These new versions may or may not introduce new documents to the pool. We observed that for $71.4\%$ of the queries there is less than $50\%$ overlap between the top ten URLs retrieved for the different quoted versions. This indicates that different ways of quoting the segments in a query does make a difference in the search results.
 By varying the pooling depth (ten in our case), one can roughly control the number of relevant and non-relevant documents entering the collection.


For each query-URL pair, where the URL has been retrieved for at least one of the quoted versions of the query (approx. $28$ per query), we obtained three independent sets of relevance judgments from human users. These users were different from annotators $H_{A}$, $H_{B}$ and $H_{C}$ who marked the segmentations, but having similar familiarity with search systems. For each query, the corresponding set of URLs was shown to the users after deduplication and randomization (to prevent position bias for top results), and asked to mark whether the URL was {\em irrelevant} (score = $0$), {\em partially relevant} (score = $1$) or {\em highly relevant} (score = $2$) to the query. We then computed the average rating for each query-URL pair (the entire set forming $\Rcal$), which has been used for subsequent nDCG, MAP and MRR computations. Please refer to Table~\ref{tab:inter_anno} in Sec.~\ref{subsec:inter} for inter-annotator agreement figures and other related discussions. 

\subsection{Segmentation algorithms}
\label{subsec:candidates}

\begin{table}	[tbp] \scriptsize
	\centering
	\caption{Segmentation algorithms compared on our framework.}
		\begin{tabular} {r l}
			\\ \toprule
			\textbf{Algorithm} 										& \textbf{Training data}												\\ \toprule
			Li et al.~\cite{li:11}								& Click data, Web $n$-gram probabilities				\\
			Hagen et al.~\cite{hagen:11}					& Web $n$-gram frequencies, Wikipedia titles		\\
			Mishra et al.~\cite{mishra:11} 				& Query logs																		\\	
			\cite{mishra:11} + Wiki								& Query logs, Wikipedia titles									\\
			PMI-W~\cite{hagen:11}									& Web $n$-gram probabilities (used as baseline)	\\
			PMI-Q~\cite{mishra:11} 								& Query logs (used as baseline)									\\
			\bottomrule
		\end{tabular}
	\label{tab:candidates}
\end{table}

Table~\ref{tab:candidates} lists the six segmentation algorithms that have been studied in this work. Li et al.~\cite{li:11} use the expectation maximization algorithm to arrive at the most probable segmentation, while Hagen et al.~\cite{hagen:11} show a simple frequency-based method produces a performance comparable to the state-of-the-art. The technique in Mishra et al.~\cite{mishra:11} uses only query logs for segmenting queries. In our experiments, we observed that the performance of Mishra et al.~\cite{mishra:11} can be improved if we used Wikipedia titles. We refer to this as ``\cite{mishra:11} + Wiki" in our experiments (see Appendix~A for details). The Point-wise Mutual Information (PMI)-based algorithms are used as baselines. The thresholds for PMI-W and PMI-Q were chosen to be 8.141 and 0.156 respectively, that maximized the \textit{Seg-F} (see Sec.~\ref{subsec:match}) on our development set.

\subsection{Public release of data}
\label{subsec:public}

The test set of search queries along with their manual and some of the algorithmic segmentations, the theoretical best segmentation output that can serve as an evaluation benchmark ($BQV_{BF}$ in Sec.~\ref{subsec:lucene}), and the list of URLs whose contents serve as our document corpus is available for public use\footnote{\scriptsize \url{http://cse.iitkgp.ac.in/resgrp/cnerg/qa/querysegmentation.html}}. The relevance judgments for the query-URL pairs have also been made public which will enable the community to use this dataset for evaluation of any new segmentation algorithm.

\section{Experiments and Observations}
\label{sec:discussion}

\begin{table*}[t] 
\centering
\caption{Results of IR-based evaluation of segmentation algorithms using Lucene (mean oracle scores).}
\newcolumntype{G}{>{\columncolor [gray] {0.90}}c}
\begin{tabular} {c G c c c c c c G G G c}
\\ \toprule
\textbf{Metric} & \textbf{Unseg.} & \textbf{\cite{li:11}} & \textbf{\cite{hagen:11}} & \textbf{\cite{mishra:11}} & \textbf{\cite{mishra:11} +} & \textbf{PMI-W} & \textbf{PMI-Q} & {\boldmath $H_{A}$} & {\boldmath $H_{B}$} & {\boldmath $H_{C}$} & {\boldmath $BQV_{BF}$} \\
& \textbf{query} & & & & \textbf{Wiki} & & & & & & \\ \toprule
\textbf{nDCG@5} & 0.688 & 0.752* & \textbf{0.763}* & 0.745 & \textbf{0.767}* & 0.691 & \textbf{0.766}* & \textbf{0.770} & 0.768 & 0.759 & 0.825\\
\textbf{nDCG@10} & 0.701 & 0.756* & \textbf{0.767}* & 0.751 & \textbf{0.768}* & 0.704 & \textbf{0.767}* & \textbf{0.770} & \textbf{0.768} & \textbf{0.763} & 0.832\\ \midrule
\textbf{MAP@5} & 0.882 & 0.930* & \textbf{0.942}* & 0.930* & \textbf{0.945}* & 0.884 & 0.932* & \textbf{0.944} & \textbf{0.942} & 0.936 & 0.958\\ 
\textbf{MAP@10} & 0.865 & 0.910* & \textbf{0.921}* & 0.910* & \textbf{0.923}* & 0.867 & 0.912* & \textbf{0.923} & \textbf{0.921} & \textbf{0.916} & 0.944\\ \midrule
\textbf{MRR@5} & 0.538 & 0.632* & \textbf{0.649}* & 0.609 & \textbf{0.650}* & 0.543 & \textbf{0.648}* & \textbf{0.656} & \textbf{0.648} & 0.632 & 0.711\\
\textbf{MRR@10} & 0.549 & 0.640* & \textbf{0.658}* & 0.619 & \textbf{0.658}* & 0.555 & \textbf{0.656}* & \textbf{0.665} & \textbf{0.656} & 0.640 & 0.717\\ \bottomrule
\end{tabular}
\scriptsize \\ The highest value in a row (excluding the $BQV_{BF}$ column) and those with no statistically significant difference with the highest value are marked in \textbf{boldface}. The values for algorithms that perform better than or have no statistically significant difference with the \textit{minimum} of the human segmentations are marked with *. The paired $t$-test was performed and the null hypothesis was rejected if the $p$-value was less than $0.05$.
\label{tab:lucene}
\end{table*}

\begin{table*}
	\centering
	\caption{Matching metrics for different segmentation algorithms and human annotations \textit{with $BQV_{BF}$ as reference.}}
	\newcolumntype{G}{>{\columncolor [gray] {0.90}}c}
		\begin{tabular}{c G c c c c c c G G G c}
			\\ \toprule
			\textbf{Metric}	& \textbf{Unseg.} & \textbf{\cite{li:11}} & \textbf{\cite{hagen:11}} 	& \textbf{\cite{mishra:11}} & \textbf{\cite{mishra:11} +} & \textbf{PMI-W} 	& \textbf{PMI-Q}	& {\boldmath $H_{A}$} & {\boldmath $H_{B}$} & {\boldmath $H_{C}$}	& {\boldmath $BQV_{BF}$}\\			
				& \textbf{query} & & & & \textbf{Wiki} & & & & & & \\ \toprule
			\textbf{Qry-Acc}	& 0.044						& 0.056	 & 0.082*	& 0.058	& 0.094* 					& 0.046						& \textbf{0.104}* & 0.086 					& 0.074 & 0.064 & 1.000 \\
			\textbf{Seg-Prec}	& \textbf{0.226}*	& 0.176* & 0.189*	&	0.206*	& 0.203*					& \textbf{0.229}*	& 0.218* 					& 0.176 					& 0.166 & 0.178	& 1.000 \\
			\textbf{Seg-Rec}	& \textbf{0.325}*	& 0.166* & 0.162*	&	0.210*	& 0.174*					& \textbf{0.323}*	& 0.196* 					& 0.144 					& 0.133 & 0.154	& 1.000 \\
			\textbf{Seg-F}		& \textbf{0.267}*	& 0.171* & 0.174*	&	0.208*		& 0.187*						& \textbf{0.268}*	& 0.206*	 					& 0.158 					& 0.148 & 0.165	& 1.000 \\
			\textbf{Seg-Acc}	& 0.470						& 0.624  & 0.661*	&	0.601		& 0.667*	& 0.474	& 0.660* 					& \textbf{0.675}	& \textbf{0.675} & 0.663	& 1.000 \\ \bottomrule
		\end{tabular}
\scriptsize \\ The highest value in a row (excluding the $BQV_{BF}$ column) and those with no statistically significant difference with the highest value are marked in \textbf{boldface}. The values for algorithms that perform better than or have no statistically significant difference with the \textit{minimum} of the human segmentations are marked with *. The paired $t$-test was performed and the null hypothesis was rejected if the $p$-value was less than $0.05$.
	\label{tab:best_ir}
\end{table*}

\begin{table*}
	\centering
	\caption{Performance of PMI-Q and~\cite{li:11} with respect to matching (mean of comparisons with $H_{A}$, $H_{B}$ and $H_{C}$ as references) and IR metrics.}
	\newcolumntype{G}{>{\columncolor [gray] {0.90}}c}
		\begin{tabular}{c G G G c c c c c}
			\\ \toprule
			\textbf{Metric}	& \textbf{nDCG@10} & \textbf{MAP@10} & \textbf{MRR@10} 	& \textbf{Qry-Acc} & \textbf{Seg-Prec} & \textbf{Seg-Rec} & \textbf{Seg-F}	& \textbf{Seg-Acc} \\ \toprule
			\textbf{PMI-Q}	& \textbf{0.767}	& \textbf{0.912} & \textbf{0.656}	& 0.341	& 0.448 & 0.487	& 0.467 & \textbf{0.810} \\
			\textbf{\cite{li:11}}	& 0.756	& 0.910 & 0.640	&	\textbf{0.375}	& \textbf{0.524}	& \textbf{0.588}	& \textbf{0.554} & \textbf{0.810} \\ \bottomrule	
		\end{tabular}
		\scriptsize \\ The highest values in a column are marked in \textbf{boldface}. 
	\label{tab:pmi_li}
\end{table*}

In this section we present experiments, results and the key inferences made from them.

\subsection{IR Experiments}
\label{subsec:lucene}

For the retrieval-based evaluation experiments, we use the Lucene\footnote{\scriptsize \url{http://lucene.apache.org/java/docs/index.html}} text retrieval system, which is publicly available as a code library. In its default configuration, Lucene does not perform any automatic query segmentation, which is very important for examining the effectiveness of segmentation algorithms in an IR-based scheme. Double quotes can be used in a query to force Lucene to match the quoted \textit{phrase} (in Lucene terms) exactly in the documents. Starting with the segmentations output by each of the six algorithms as well as the three human annotations, we generated all possible quoted query versions, which resulted in a total of $4,746$ versions for the $500$ queries. In the notation of Sec.~\ref{sec:evaluation}, this corresponds to generating $\Qcal_\Acal(\qbf)$ for each segmentation method $\Acal$ (including one for each human segmentation) and for every query $\qbf\in\Qcal$. These quoted versions were then passed through Lucene to retrieve documents from the pool. For each segmentation scheme, we then use the oracle described in Sec.~\ref{sec:evaluation} to obtain the query version yielding the best result (as determined by the IR metrics -- nDCG, MAP and MRR computed according to the human relevance judgments). These oracle scores are then averaged over the query set to give us the QVRS measures.

The results are summarized in Table~\ref{tab:lucene}. Different rows represent the different IR metrics that were used and columns correspond to different segmentation strategies. The second column (marked ``Unseg. Query") refers to the original unsegmented query. This can be assumed to be generated by a trivial segmentation strategy where each word is always a separate segment. Columns~3-8 denote the six different segmentation algorithms and 9-11 (marked $H_{A}$, $H_{B}$ and $H_{C}$) represent the human segmentations. The last column represents the performance of the \textbf{b}est \textbf{q}uoted \textbf{v}ersions (denoted by $BQV_{BF}$ in table) of the queries which are computed by \textit{\textbf{b}rute \textbf{f}orce}, i.e. an exhaustive search over all possible ways of quoting the parts of a query ($2^{l-1}$ possible quoted versions for an $l$-word query) irrespective of any segmentation algorithm. The results are reported for two sizes of retrieved URL lists ($k$), namely five and ten. Since we needed to convert our graded relevance judgments to binary values for computing MAP@\textit{k}, URLs with ratings of $1$ and $2$ were considered as relevant (responsible for the generally high values) and those with $0$ as irrelevant. For MRR, only URLs with ratings of $2$ were considered as relevant.


The first observation we make from the results is that human as well as all algorithmic segmentation schemes consistently outperform unsegmented queries for all IR metrics. Second, we observe that the performance of some segmentation algorithms are comparable and sometime even marginally better than some of the human annotators. Finally, we observe that there is considerable scope for improving IR performance through better segmentation (all values less than $BQV_{BF}$). The inferences from these observations are stated later in this section.

\subsection{Performance under traditional matching metrics}
\label{subsec:match}

In the next set of experiments we study the utility of traditional matching metrics that are used to evaluate query segmentation algorithms against a gold standard of human segmented queries (henceforth referred to as the {\em reference} segmentation). These metrics are listed below~\cite{hagen:11}:
\begin{enumerate}
	\item \textbf{Query accuracy (\textit{Qry-Acc}):} The fraction of queries where the output matches exactly with the reference segmentation.
	\item \textbf{Segment precision (\textit{Seg-Prec}):} The ratio of the number of segments that overlap in the output and reference segmentations to the number of output segments, averaged across all queries in the test set.
	\item \textbf{Segment recall (\textit{Seg-Rec}):} The ratio of the number of segments that overlap in the output and reference segmentations to the number of reference segments, averaged across all queries in the test set.
	\item \textbf{Segment F-score (\textit{Seg-F}):} The harmonic mean of \textit{Seg-Prec} and \textit{Seg-Rec}.
	\item \textbf{Segmentation accuracy (\textit{Seg-Acc}):} The ratio of correctly predicted boundaries and non-boundaries in the output segmentation with respect to the reference, averaged across all queries in the test set.
\end{enumerate}

We computed the matching metrics for various segmentation algorithms against $H_A$, $H_B$ and $H_C$. According to these metrics, ``Mishra et al.~\cite{mishra:11} + Wiki" turns out to be the best algorithm which agrees with the results of IR evaluation. However, the average Kendall-Tau rank correlation coefficient\footnote{\scriptsize This coefficient is $1$ when there is perfect concordance between the rankings, and $-1$ if the trends are reversed.} between the ranks of the strategies as obtained from the IR metrics (Table~\ref{tab:lucene}) and the matching metrics was only $0.75$. This indicates that matching metrics are not perfect predictors for IR performance. In fact, we discovered some costly flaws in the relative ranking produced by matching metrics. One such case was rank inversions between Li et al.~\cite{li:11} and PMI-Q. The relevant results are shown in Table~\ref{tab:pmi_li}, which demonstrate that while PMI-Q consistently performs better than Li et al.~\cite{li:11} under IR-based measures, the opposite inference would have been drawn if we had used any of the matching metrics.

In Bergsma and Wang~\cite{bergsma:07}, human annotators were asked to segment queries such that segments matched exactly in the relevant documents. This essentially corresponds to determining the best quoted versions for the query. Thus, it would be interesting to study how traditional matching metrics would perform if the humans actually marked the best quoted versions. In order to evaluate this, we used the matching metrics to compare the segmentation outputs by the algorithms and human annotations against $BQV_{BF}$. The corresponding results are quoted in Table~\ref{tab:best_ir}. The results show that matching metrics are very poor indicators of IR performance with respect to the $BQV_{BF}$. For example, for three out of the five matching metrics, the unsegmented query is ranked the best. This shows that even if human annotators managed to correctly guess the best quoted versions, the matching metrics would fail to estimate the correct relative rankings of the segmentation algorithms with respect to IR performance. This fact is also borne out in the Kendall-Tau rank correlation coefficients reported in Table~\ref{tab:kendall}. Another interesting observation from these experiments is that \textit{Seg-Acc} emerges as the best matching metric with respect to IR performance, although its correlation coefficient is still much below one.

\begin{table} [ht]
	\centering
	\caption{Kendall-Tau coefficients between IR and matching metrics \textit{with {\boldmath $BQV_{BF}$} as reference for the latter}.}
	\resizebox{0.45\textwidth}{!}{
		\begin{tabular}{c c c c c c}
		\\ \toprule
			\textbf{Metric} 	& \textbf{Qry-Acc} 	& \textbf{Seg-Prec} & \textbf{Seg-Rec} 	& \textbf{Seg-F} 	& \textbf{Seg-Acc}	\\ \toprule
			\textbf{nDCG@10} 	& 0.432 						& -0.854 						& -0.886 						& -0.854 					& \textbf{0.674} 		\\
			\textbf{MAP@10} 	& 0.322 						& -0.887 						& -0.920 						& -0.887 					& \textbf{0.750}		\\
			\textbf{MRR@10} 	& 0.395 						& -0.782 						& -0.814 						& -0.782 					& \textbf{0.598}		\\	\bottomrule		
		\end{tabular}}
		\scriptsize \\ The highest value in a row is marked in \textbf{boldface}.
	\label{tab:kendall}
\end{table}

\subsection{Inferences}
\label{subsec:insights}

{\bf Segmentation is helpful for IR.} By definition, $\Omega_\cdot(\cdot,\cdot)$ (i.e., the oracle) values for every IR metric for any segmentation scheme are at least as large as the corresponding values for the unsegmented query. Nevertheless, for every IR metrics, we observe significant performance benefits for all the human and algorithmic segmentations (except for PMI-W) over the unsegmented query. This indicates that segmentation is indeed helpful for boosting IR performance. Thus, our results validate the prevailing notion and some of the earlier observations~\cite{bendersky:09,li:11} that segmentation can help improve IR.

\textbf{Human segmentations are a good proxy, but not a true gold standard.} Our results indicate that human segmentations perform reasonably well in IR metrics. The best of the human annotators beats all the segmentation algorithms, on almost all the metrics. Therefore, evaluation against human annotations can indeed be considered as the second best alternative to an IR-based evaluation (though see below for criticisms of current matching metrics). However, if the objective is to improve IR performance, then human annotations cannot be considered a true gold standard. There are at least three reasons for this:

First, in terms of IR metrics, some of the state-of-the-art segmentation algorithms are performing as well as human segmentations (no statistically significant difference). Thus, further optimization of the matching metrics against human annotations is not going to improve the IR performance of the segmentation algorithms. Thus, evaluation on human annotations might become a limiting factor for the current segmentation algorithms.

Second, the IR performance of the best quoted version of the queries derived through our framework is significantly better than that of human annotations (last column, Table~\ref{tab:lucene}). This means that humans fail to predict the correct boundaries in many instances. Thus, there is scope for improvement for human annotations.

Third, IR performance of at least one of the three human annotators ($H_C$) is worse than some of the algorithms studied. In other words, while some annotators (such as $H_A$) are good at guessing the ``correct" segment boundaries that will help IR, not all annotators can do it well. Therefore, unless the annotators are chosen and guided properly, one cannot guarantee the quality of annotated data for query segmentation. If the queries in the test set have multiple intents, this issue becomes an even bigger concern.

\textbf{Matching metrics are misleading.} As discussed earlier and demonstrated by Tables~\ref{tab:best_ir} and \ref{tab:kendall}, the matching metrics provide unreliable ranking of the segmentation algorithms even when applied against a true gold standard, $BQV_{BF}$, that by definition maximizes IR performance. This counter-intuitive observation can be explained in two ways. Either the matching metrics or the IR metrics (or probably both) are misleading. Given that IR metrics are well-tested and generally assumed to be acceptable, we are forced to conclude that the matching metrics do not really reflect the quality of a segmentation with respect to a gold standard. Indeed, this can be illustrated by a simple example.

\textbf{\emph{Example.}}	Let us consider the query {\tt the looney toons show cartoon network}, whose best quoted version turns out to be {\tt "the looney toons show" "cartoon network"}. The underlying segmentation that can give rise to this and therefore can be assumed to be the reference is:\\
\hspace{0.5cm} Ref: {\tt the looney toons show | cartoon network}\\
The segmentations \\
\hspace{1.5cm} (1) {\tt the looney | toons show | cartoon | network}\\
\hspace{1.5cm} (2) {\tt the | looney | toons show cartoon | network}\\
are  equally bad if one considers the matching metrics of \textit{Qry-Acc}, \textit{Seg-Prec}, \textit{Seg-Rec} and \textit{Seg-F} (all values being zero) with respect to the reference segmentation. \textit{Seg-Acc} values for the two segmentations are $3/5$ and $1/5$ respectively. However, the BQV for (1) ({\tt "the looney" "toons show" cartoon network}) fetches better pages than the BQV of (2) ({\tt the looney toons show cartoon network}). So the segmentation (2) provides no IR benefit over the unsegmented query and hence performs worse than (1) on IR metrics. However, the matching metrics, except for \textit{Seg-Acc} to some extent, fail to capture this difference between the segmentations.

\begin{figure} 
	\centering
		\includegraphics[width=0.9\columnwidth]{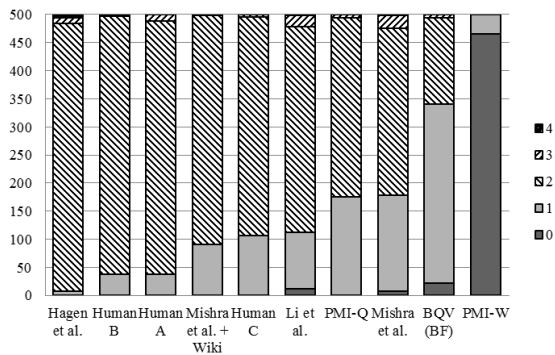}
	\caption{Distribution of multiword segments in queries across segmentation strategies.}
	\label{fig:multiword}
\end{figure}

\textbf{Distribution of multiword segments across queries gives insights about effectiveness of strategy.} The limitation of the matching metrics can also be understood from the following analysis of the multiword segments in the queries. Fig.~\ref{fig:multiword} shows the distribution of queries having a specific number of multiword segments (for example, $1$ in the legend indicates the proportion of queries having \textit{one} multiword segment) when segmented according to the various strategies. We note that for Hagen et al.~\cite{hagen:11}, $H_{B}$, $H_{A}$ and ``Mishra et al.~\cite{mishra:11} + Wiki", almost all of the queries have two multiword segments. For $H_{C}$, Li et al.~\cite{li:11},  PMI-Q and Mishra et al.~\cite{mishra:11}, the proportion of queries that have only one multiword segment increases. Finally, PMI-W has almost negligible queries with a multiword segment. $BQV_{BF}$ is different from all of them and has a majority of queries with one multiword segment. Now given that the first group generally does the best in IR, followed by the second, we can say that out of the two multiword segments marked by these strategies, only one needs to be quoted. PMI-W as well as unsegmented queries are bad because these schemes cannot detect the one crucial multiword segment quoting which improves the performance. Nevertheless, these schemes do well for matching metrics against $BQV_{BF}$ because both have a large number of single word segments. Clearly this is not helpful for IR. Finally, Mishra et al.~\cite{mishra:11} performs poorly despite being able to identify a multiword segment in most of the cases because it is not identifying the one that is important for IR.

Hence, the matching metrics are misleading due to two reasons. First, they do not take into account that splitting a useful segment (i.e., a segment which should be quoted to improve IR performance) is less harmful than joining two unrelated segments. Second, matching metrics are, by definition, agnostic to which segments are useful for IR. Therefore, they might unnecessarily penalize a segmentation for not agreeing on the segments which should not be quoted, but are present in the reference human segmentation. While the latter is an inherent problem with any evaluation against manually segmented datasets, the former can be resolved by introducing a new matching metric that differentially penalizes splitting and joining of segments. This is an important and interesting research problem that we would like to address in the future. However, we would like to emphasize here that with the IR system expected to grow in complexity in the future (supporting more flexible matching criteria), the need for an IR-based evaluation like ours' becomes imperative.

Based on our new evaluation framework and corresponding experiments, we observe that ``Mishra et al.~\cite{mishra:11} + Wiki" has the best performance. Nevertheless, the algorithms are trained and tested on different datasets, and therefore, a comparison amongst the algorithms might not be entirely fair. This is not a drawback of the framework and can be circumvented by appropriately tuning all the algorithms on similar datasets. However, the objective of the current work is not to compare segmentation algorithms; rather, it is to introduce the evaluation framework, gain insights from the experiments and highlight the drawbacks of human segmentation-based evaluation.

\section{Related issues}
\label{sec:relatedis}

In this section, we will briefly discuss a few related issues that are essential for understanding certain design choices and decisions made during the course of this research. 

\subsection{Motivation for a new dataset}
\label{subsec:newdata}

TREC data has been a popular choice for conducting IR-based experiments throughout the past decade. Since there is no track specifically geared towards query segmentation, the queries and \textit{qrels} (query-relevance sets) from the ad hoc retrieval task for the Web Track would seem the most relevant to our work. However, $74\%$ of the $50$ queries in the $2010$ Web track ad hoc task had less than three words. Also, when these $50$ queries were segmented using the six algorithms, half of the queries did not have a multiword segment. As discussed earlier, query segmentation is useful but not necessarily for all types of queries. The benefit of segmentation may be observed only when there are multiple multiword segments in the queries. The TREC Million Query Track, last held in $2009$, has a much larger set of $40,000$ queries, with a better coverage of longer queries. But since the goal of the track is to test the hypothesis that a test collection built from several incompletely judged topics is a better tool than a collection built using traditional TREC pooling, there are only about $35,000$ query-document relevance judgments for the $40,000$ queries. Such a sparse \textit{qrels} is not suitable here -- incomplete assessments, especially for documents near the top ranks, could cause crucial errors in system comparisons. Yet another option could have been to use BWC07 as \Qcal  and create the corresponding \Ucal  and  \Rcal. However, this query set is known to suffer from several drawbacks~\cite{hagen:11}. A new dataset for query segmentation\footnote{\scriptsize \url{http://bit.ly/xIhSur}} containing manual segment markups collected through crowdsourcing has been recently made publicly available (after we had completed construction of our set) by 
Hagen et al.~\cite{hagen:11}, but it lacks query-document relevance judgments. These factors motivated us to create a new dataset suitable for our framework, which has been made publicly available (see Sec.~\ref{subsec:public}).

\subsection{Retrieval using Bing}
\label{subsec:bing}

\begin{table}[t]	\scriptsize
	\centering
	\caption{IR-based evaluation using Bing API.}
	\newcolumntype{G}{>{\columncolor [gray] {0.90}}c}
		\begin{tabular} {c G c c G G c c}
		\\ \toprule
			\textbf{Metric}		& \textbf{Unseg.}	& \textbf{All quoted for}						& \textbf{Oracle for} 							\\
												& \textbf{query}	& \textbf{\cite{mishra:11} + Wiki}	& \textbf{\cite{mishra:11} + Wiki} 	\\ \toprule
			\textbf{nDCG@10}	& 0.882 					& 0.823 														& \textbf{0.989*} 									\\ 
			\textbf{MAP@10} 	& 0.366 					& 0.352 														& \textbf{0.410*} 									\\ 
			\textbf{MRR@10} 	& 0.541 					& 0.515 														& \textbf{0.572*} 									\\ \bottomrule
		\end{tabular}
		\scriptsize \\ The highest value in a row is marked \textbf{bold}. Statistically significant ($p$ < 0.05 for paired $t$-test) improvement over the unsegmented query is marked with *.
	\label{tab:bing_api}
\end{table}

Bing is a large-scale commercial Web search engine that provides an API service. Instead of Lucene, which is too simplistic, we could have used Bing as the IR engine in our framework. However, such a choice suffers from two drawbacks. First, Bing might already be segmenting the query with its own algorithm as a preprocessing step. Second, there is a serious replicability issue. The document pool that Bing uses, i.e. the Web, changes dynamically with documents added and removed from the pool on a regular basis. This makes it difficult to publish a static gold standard dataset with relevance judgments for all appropriate query-URL pairs that the Bing API may retrieve even for the same set of queries. In view of this, the main results were reported in this paper using the Lucene text retrieval system.

However, since we used Bing API to construct \Ucal and corresponding \Rcal , we have the evaluation statistics using the Bing API as well. For paucity of space, in Table~\ref{tab:bing_api} we only present the results for nDCG@10, MRR@10 and MAP@10 for ``Mishra et al.~\cite{mishra:11} + Wiki". The table reports results for three quoted version-selection strategies: (i) Unsegmented query only (equivalent to each word being within quotes) (ii) All segments quoted and (iii) {\em QVRS} (oracle for ``Mishra et al.~\cite{mishra:11} + Wiki"). For all the three metrics, {\em QVRS} is statistically significantly higher than results for the unsegmented query. Thus, segmentation can play an important role towards improving IR performance of the search engine. We note that the strategy of quoting all the segments is, in fact, detrimental to IR performance. This emphasizes the point that how the segments should be matched in the documents is a very important research challenge. Instead of quoting all the segments, our proposal here is to assume an oracle that will suggest which segments to quote and which are to be left unquoted for the best IR performance. Philosophically, this is a major departure from the previous ideas of using quoted segments, because re-issuing a query by quoting all the segments implies segmentation as a way to generate a fully quoted version of the query (all segments in double quotes). This definition severely limits the scope of segmentation, which ideally should be thought of as a step forward better query understanding.

\begin{table}[t]
	\centering
	\caption{Inter-annotator agreement on features as observed from our experiments.}
		\begin{tabular} {c c c c c}
			\\ \toprule
			\textbf{Feature} 		& \textbf{Pair 1} & \textbf{Pair 2} & \textbf{Pair 3}	& \textbf{Mean} \\ \toprule
			\textbf{Qry-Acc} 		& 0.728 					& 0.644 					& 0.534 					& 0.635 				\\
			\textbf{Seg-Prec} 	& 0.750 					& 0.732 					& 0.632 					& 0.705 				\\
			\textbf{Seg-Rec} 		& 0.756 					& 0.775 					& 0.671 					& 0.734 				\\
			\textbf{Seg-F} 			& 0.753 					& 0.753 					& 0.651 					& 0.719 				\\
			\textbf{Seg-Acc}		&	0.911 					& 0.914 					& 0.872 					& 0.899 				\\ \midrule
			\textbf{Rel. judg.} & 0.962 					& 0.959 					& 0.969 					& 0.963 				\\ \bottomrule
		\end{tabular}
		\\ \scriptsize For relevance judgments, only pairs of (0, 2) and (2, 0) were considered disagreements.
	\label{tab:inter_anno}
\end{table}

\subsection{Inter-annotator agreement}
\label{subsec:inter}

Inter-annotator agreement (IAA) is an important indicator for reliability of manually created data. Table~\ref{tab:inter_anno} reports the pairwise IAA statistics for $H_{A}$, $H_{B}$ and $H_{C}$. Since there are no universally accepted metrics for IAA, we report the values of the five matching metrics when one of the annotations (say $H_{A}$) is assumed to be the reference and the remaining pair ($H_{B}$ and $H_{C}$) is evaluated against it (average reported). As is evident from the table, the values of all the metrics, except for {\em Seg-Acc}, is less than $0.78$ (similar values reported in~\cite{tan:08}), which indicates a rather low IAA. The value for \textit{Seg-Acc} is close to $0.9$, which to the contrary, indicates reasonably high IAA (as in~\cite{tan:08}). The last row of Table~\ref{tab:inter_anno} reports the IAA for the three sets of relevance judgments (therefore, the actual pairs for this column are different from that of the other rows). The agreement in this case is quite high.

There might be several reasons for low IAA for segmentation, such as lack of proper guidelines and/or an inherent inability of human annotators to mark the correct segments of a query. Low IAA raises serious doubts about the reliability of human annotations for query segmentation. On the other hand, high IAA for relevance judgments naturally makes these annotations much more reliable for any evaluation, and strengthens the case for our IR-based evaluation framework which only relies on relevance judgments. We note that ideally, relevance judgments should be obtained from the user who has issued the query. This has been referred to as {\em gold} annotations, as opposed to {\em silver} or {\em bronze} annotations which are obtained from expert and non-expert annotators respectively who have not issued the query~\cite{bailey:08}. Gold annotations are preferable over silver or bronze ones due to relatively higher IAA. Our annotations are silver standard, though very high IAA essentially indicates that they might be as reliable as gold standard. The high IAA might be due to the unambiguous nature of the queries.

\section{Related work}
\label{sec:relatedwk}

Since its inception in 2003~\cite{risvik:03}, many algorithms have been proposed for automatic segmentation of Web queries. The approaches vary from purely supervised~\cite{bergsma:07} to fully unsupervised~\cite{hagen:11,mishra:11} machine learning techniques. They differ widely in terms of resources usage (Table~\ref{tab:candidates}) and the underlying algorithmic techniques (e.g., expectation maximization~\cite{tan:08} and eigenspace similarity~\cite{zhang:09}).

\subsection{Evaluation on manual annotations}
\label{subsec:manual}

Despite the diversity in approaches to the task, till date there has been only one standard approach for evaluation of query segmentation algorithms, which is to compare the machine output against a set of queries segmented by humans~\cite{bergsma:07,brenes:10,hagen:11,li:11,mishra:11,tan:08,zhang:09}. The basic assumption underlying this evaluation scheme is that humans are capable of segmenting a query in a ``correct" or ``the best possible" way, which, if exploited appropriately, will result in maximum benefits in IR performance. This is probably motivated by the extensive use of human judgments and annotations as the gold standard in the field of NLP (e.g., parts-of-speech labeling, phrase boundary identification, etc.). However, this idea has several shortcomings, as pointed out in Sec.~\ref{subsec:insights}. Among those who validate query segmentation against human-labeled data, most~\cite{bergsma:07,brenes:10,hagen:10,hagen:11,li:11,tan:08,zhang:09} report accuracies on BWC07~\cite{bergsma:07}. The popularity of the BWC07 dataset is partly because it was one of the first human annotated datasets created for query segmentation, and partly because it is the only publicly available dataset of its kind. While BWC07 has provided a common benchmark for comparing various query segmentation algorithms, there are several limitations of this specific dataset. BWC07 only contains noun phrase queries and there is a non-trivial amount of noise in the annotations. See~\cite{hagen:11} for a detailed criticism of this dataset. 


\subsection{IR-based evaluation}
\label{subsec:ir}

There has been only a handful of studies that explore some initial ideas about IR-based evaluation~\cite{bendersky:09,hagen:11,li:11} for query segmentation. Bendersky et al.~\cite{bendersky:09} were the first to study the effects of segmentation from an IR perspective. They wanted to see if retrieval quality could be improved by incorporating knowledge of query chunks into an MRF-based retrieval system~\cite{metzler:05}. Their experiments on different TREC collections using popular IR metrics like MAP indicate that query segmentation can indeed boost IR performance. Li et al.~\cite{li:11} examined the usefulness of query segmentation when built into language models for retrieval, in a Web search setting. However, none of these studies propose an objective IR-based \textit{evaluation framework} for query segmentation. Their scope is limited to the demonstration of one particular strategy for exploiting segmentations for improving IR, instead of evaluating and comparing a set of algorithms. 

As an excursus to their main work, Hagen et al.~\cite{hagen:11} examined if submitting fully quoted queries (generated from algorithm outputs) results in fetching better pages by the search engines. They study the top fifty retrieved documents when the following versions of the queries -- unsegmented, manually quoted, quoted by the technique in Bergsma and Wang~\cite{bergsma:07}, and by their own method -- are submitted to Bing. Assuming the pages retrieved by manual quotation as relevant, it was observed that the technique in Bergsma and Wang~\cite{bergsma:07} achieves the highest average recall. However, the authors also state that such an assumption need not hold good in reality and emphasized the need for an in-depth retrieval-based evaluation.


We would like to emphasize here that the aim of a segmentation technique is not to come up with the best quoted version of a query. While some past works have explicitly or implicitly assumed this definition, there are also other works that view segmentation as a purely structural analysis of a query that identifies chunks or sequences of words that are semantically connected as a unit~\cite{li:11,mishra:11}. By quoting all the segments we would be penalizing the latter philosophy of segmentation, which is a  more productive and practically useful view.

There have been a few studies on detection of noun phrases from queries~\cite{carvalho:10,zhang:07}. This task is similar to query segmentation in the sense that the phrase can be considered as a single unit in the query. Zhang et al.~\cite{zhang:07} has shown that such phrase detection schemes can actually help in retrieval, and therefore, is along the lines of the philosophy of the present evaluation framework. Nevertheless, as far as we know, this is the first time that a formal conceptual framework for an IR-based evaluation of query segmentation has been proposed. Our study, also for the first time, compares the effectiveness of human segmentation and related matching metrics to an IR-based evaluation. 

\section{Conclusions and future work}
\label{sec:conclusions}

End-user of query segmentation is the retrieval engine; hence, it is essential that any segmentation algorithm should be evaluated in an IR-based framework. In this research, we overcome several conceptual challenges to design and implement the first such scheme of evaluation for query segmentation. Using a carefully selected query test set and a group of segmentation strategies, we show that it is possible to have a fair comparison of the relative goodness of each strategy as measured by standard IR metrics. The proposed framework uses resources which are essential for any IR system evaluation, and hence does not require any special input. Our entire dataset -- complete with queries, segmentation outputs and relevance judgments -- has also been made publicly available to facilitate further research by the community.
 
Moreover, we gain several useful and non-intuitive insights from the evaluation experiments.  Most importantly, we show that human notions of query segments may not be the best for maximizing retrieval performance, and treating them as the gold standard limits the scope for improvement for an algorithm. Also, the matching metrics extensively used till date for comparing against gold standard segmentations can often be misleading. We would like to emphasize that in the future, the focus of IR will mostly shift to tail queries. In such a scenario, an IR-based evaluation scheme gains relevance because validation against a fixed set of gold standard segmentation may often lead to overfitting of the algorithms without yielding any real benefit.
 
A hypothetical oracle has been shown to be quite useful, but we realize that it will  be a much bigger contribution to the community if we could implement a context-aware oracle that can actually tell the search engine which version of a segmented query should be chosen at runtime.
\section{Acknowledgments}
\label{sec:acknowledgments}

We would like to thank Bo-June (Paul) Hsu and Kuansan Wang (Microsoft Research, Redmond), for providing us with the code for Li et al.~\cite{li:11}. We also thank Matthias Hagen (Bauhaus Universit\"{a}t Weimar), for providing us with the segmentation output of Hagen et al.~\cite{hagen:11} on our test set at a very short notice. The first author was supported by Microsoft Corporation and Microsoft Research India under the Microsoft Research India PhD Fellowship Award.

\bibliographystyle{abbrv}
\bibliography{rishiraj}

\section*{APPENDIX A: WIKI-BOOST}
\label{sec:appsec}

\begin{algorithm}
	\label{alg:wiki}
	\caption{Wiki-Boost($Q'$, $W$)}	
	\begin{algorithmic} [1]
		\STATE $W' \gets \emptyset$
		\FORALL{$w \in W$}
			\STATE $w' \gets Seg\mathchar`\-Phase\mathchar`\-1(w)$
			\STATE $W' \gets W' \cup w'$
		\ENDFOR
		\STATE $W'\mathchar`\-scores \gets \emptyset$
		\FORALL{$w' \in W'$}
			\STATE $w'\mathchar`\-score \gets PMI(w')\;based\;on\;Q'$
			\STATE $W'\mathchar`\-scores \gets W'\mathchar`\-scores \cup w'\mathchar`\-score$
		\ENDFOR
		\STATE $U\mathchar`\-scores \gets \emptyset$
		\FORALL{$unique\;unigrams\;u \in Q'$}
			\STATE $u\mathchar`\-score \gets probability(u)\;in\;Q'$
			\STATE $U\mathchar`\-scores \gets U\mathchar`\-scores \cup u\mathchar`\-score$
		\ENDFOR
		\STATE $W'\mathchar`\-scores \gets W'\mathchar`\-scores \cup U\mathchar`\-scores$
		\RETURN $W'\mathchar`\-scores$
	\end{algorithmic}
\end{algorithm}

In this appendix, we explain how to augment the output of an $n$-gram score aggregation based segmentation algorithm with Wikipedia titles\footnote{\scriptsize \url{http://dumps.wikimedia.org/enwiki/latest/}, accessed April 6, 2011}. Input to \textit{Wiki-Boost} is a list of queries $Q'$ already segmented by the algorithm in Mishra et al.~\cite{mishra:11} (or any algorithm that meets the above criterion) (say, \textit{Seg-Phase-}1) and $W$, the list of all stemmed Wikipedia titles ($4,508,386$ entries after removing one-word entries and those with non-ASCII characters). We compute the PMI-score of an $n$-segment Wikipedia title $w'$ (segmented by \textit{Seg-Phase-}1) by taking the higher of the PMI scores of the first $(n - 1)$ segments with the last segment \textit{and} the first segment and the last $(n - 1)$ segments. The frequencies of all $n$-grams are computed from $Q'$. Scores for unigrams are defined to be their probabilities of occurrence. Thus, the output of the \textit{Wiki-Boost} is a list of PMI-scores for each Wikipedia title in $W$.

Following this, we use a second segmentation strategy (say, \textit{Seg-Phase-}2) that takes as input $q'$ (the query $q$ segmented by \textit{Seg-Phase-}1) and tries to further join the segments of $q'$ such that the product of scores of the candidate output segments, computed based on the output of \textit{Wiki-Boost}, is maximized. A dynamic programming approach is found to be helpful in searching over all possible segmentations in \textit{Seg-Phase-}2. The output of \textit{Seg-Phase-}2 is the final segmentation output.

\end{document}